# The origin of Pluto's orbit: implications for the Solar System beyond Neptune

Renu Malhotra

*Lunar and Planetary Institute, 3600 Bay Area Blvd, Houston, TX 77058.*
*Electronic mail: renu@lpi.jsc.nasa.gov*

## ABSTRACT

*The origin of the highly eccentric, inclined, and resonance-locked orbit of Pluto has long been a puzzle. A possible explanation has been proposed recently [Malhotra, R., Nature 365:819-21 (1993)] which suggests that these extraordinary orbital properties may be a natural consequence of the formation and early dynamical evolution of the outer Solar system. A resonance capture mechanism is possible during the clearing of the residual planetesimal debris and the formation of the Oort Cloud of comets by planetesimal mass loss from the vicinity of the giant planets. If this mechanism were in operation during the early history of the planetary system, the entire region between the orbit of Neptune and approximately 50 AU would have been swept by first order mean motion resonances. Thus, resonance capture could occur not only for Pluto, but quite generally for other trans-Neptunian small bodies. Some consequences of this evolution for the present-day dynamical structure of the trans-Neptunian region are: (i) most of the objects in the region beyond Neptune and up to $\sim 50$ AU exist in very narrow zones located at orbital resonances with Neptune (particularly the 3:2 and the 2:1 resonances), and (ii) these resonant objects would have significantly large eccentricities. The distribution of objects in the Kuiper Belt as predicted by this theory is presented here.*

*Subject headings:* solar system: origins, dynamics — Pluto — comets — Kuiper Belt

# 1. Introduction

In the widely accepted paradigm for the formation of the Solar system, the planets accumulated in a highly dissipative disk of dust and gas orbiting the protosun, and most planets formed in near-circular and nearly co-planar orbits. The outermost planet, Pluto, is an oddity in this scheme; its orbit is highly eccentric (e = 0.25) and inclined (17 degrees to the ecliptic). The large eccentricity means that Pluto crosses the orbit of Neptune, and it traverses a very large region of space from just inside the orbit of Neptune at 30 AU to almost 50 AU. Indeed, soon after Pluto was first discovered in 1930, it was realized that the dynamical lifetime of this new planet was short before a close encounter with Neptune radically altered its orbit (Lyttleton 1936). Three decades later, it was found that a dynamical protection mechanism exists that prevents close encounters between Pluto and Neptune: a 120,000 year orbit integration of the outer planets by Cohen & Hubbard 1965 showed that Pluto is locked in a 3:2 orbital resonance with Neptune which maintains a large longitude separation between the planets at orbit crossing and causes Pluto's perihelion to librate about a center $\pm 90°$ away from Neptune. Since the Cohen and Hubbard work, orbit integrations of increasingly longer times have uncovered several other resonances and near-resonances in Pluto's motion (Williams & Benson 1971, Applegate *et al.* 1986, Sussman & Wisdom 1988, Milani *et al.* 1989). Perhaps the most important of these "weaker" resonances is the "argument-of-perihelion libration" which ensures that at perihelion Pluto is close to its maximum excursion above the mean plane of the Solar system; this has the effect of increasing the minimum approach distance between Pluto and Neptune and between Pluto and Uranus than would otherwise be the case. (It is worth emphasizing that the longitude-of-perihelion libration centered at $\pm 90°$ relative to Neptune and the argument-of-perihelion libration about $90°$ are two quite distinct phenomena; the latter is associated directly with the 3:2 orbital period resonance lock between Neptune and Pluto.) Pluto's orbit is confined in a very narrow region of relative orbital stability near the 3:2 Neptune resonance, a region in phase space that is bounded by highly chaotic orbits. (See Malhotra & Williams 1994 for a recent review of Pluto's orbital dynamics).

There has, of course, been much speculation as to the origin of this extraordinary orbit of Pluto (Lyttleton 1936, Harrington & Flandern 1979, Farinella *et al.* 1979, Dormand & Woolfson 1980, Olsson-Steel 1988, Malhotra 1993a, Levison & Stern 1994), all but one – Malhotra 1993a [hereafter Paper I] – of these speculations requiring one or more low-probability "catastrophic" events. A popular theme in the earlier scenarios was that Pluto is an escaped satellite of Neptune. On the other hand, studies and modeling of the physical characteristics and composition of Pluto have led to the conclusion that it accumulated in a heliocentric orbit of ice-rich planetesimals in the outer reaches of the Solar Nebula rather than in a circumplanetary disk (McKinnon & Mueller 1988, Tancredi & Fernandez 1991). In consonance with this, the recent theories by Malhotra 1993a and Levison & Stern 1994 both propose that Pluto formed in a near-circular coplanar heliocentric orbit beyond the orbits of the giant planets, but they differ in the physical and dynamical mechanisms that placed Pluto in its unusual orbit.

In the model proposed by Levison & Stern, the orbital configuration of the planets is taken as observed today, except that a "test-Pluto" is placed in an initially low-eccentricity, low-inclination orbit near the 3:2 Neptune resonance. With some fine-tuning of initial conditions, orbit integrations show that such an orbit has its eccentricity and inclination pumped up to values comparable to those of the real Pluto in a timescale of about $10^7$ years. However, the orbit remains chaotic during this evolution; Levison & Stern then propose that Pluto was "knocked" into the stable 3:2 resonance libration region by one or more dissipative collisions with a neighboring small body or bodies.

Malhotra's theory [Paper I] does not invoke any catastrophic collisions, and is possibly compatible with the standard paradigm for planet formation. (The reader is referred to Levy & Lunine 1993 for reviews of planet formation theory.) In this model, an initially low-inclination, nearly circular orbit of Pluto beyond the orbits of the giant planets evolves into its Neptune-crossing but resonance-protected orbit as a result of early dynamical evolution of the outer solar system. The physical causes of this evolution lie in the late stages of planet formation when the gravitational scattering and eventual clearing of remnant planetesimal debris by the giant planets (and the concomitant exchange of energy and angular momentum between the planets and the planetesimals) may have caused a significant evolution of the giant planet orbits (Fer-nandez & Ip 1984). In particular, Neptune's orbit may have expanded considerably, and its exterior orbital resonances would have swept through a large region of trans-Neptunian space. During this resonance sweeping, Pluto could have been captured into the 3:2 orbital period resonance with Neptune and its eccentricity (as well as inclination — see section 4) would have been pumped up during the subsequent evolution.

The dynamical mechanisms invoked in this theory are quite general, and would apply not only to the evolution of the trans-Neptunian body labeled "Pluto", but also to any other members of the trans-Neptune region. While this possibility was implicit in Paper I, it is my purpose in the present paper to make explicit the implications and predictions of this "resonance capture theory" of the origin of Pluto's orbit for the present-day architecture of the Solar system beyond Neptune.

That the outermost parts of the Solar system may be populated by primordial icy planetesimals has been conjectured on both theoretical and observational grounds. For example, Kuiper 1951 suggested this on the basis of theoretical considerations of the genesis of the planetary system from the primordial Solar Nebula. Whipple 1964 and Bailey 1983) speculated on a massive comet belt as the source of unexplained perturbations of Neptune's orbit (although this argument must now be discarded as the post-*Voyager* revisions in the planetary ephemeris no longer show any unexplained residuals in Neptune's motion (Standish 1993)). Hamid *et al.* 1968 analyzed the orbital plane perturbations of comet P/Halley and concluded that any comet belt between 40 AU and 50 AU has a total mass less than $1 M_\oplus$. More recently, it has been suggested that the observed short-period comets with orbital periods $\lesssim 20$ yr originate in a belt of low-inclination bodies just beyond the orbit of Neptune, between 35 AU and 50 AU (Fernandez 1980, Fernandez & Ip 1983). The older hypothesis that short-period comets originate in a population of near-parabolic Oort Cloud comets (which are perturbed into shorter orbits by the giant planets) appears unlikely: Duncan *et al.* 1988, Quinn *et al.* 1990 have shown that the orbital element distribution of the observed short-period comets is inconsistent with a source in the nearly isotropic Oort Cloud but is compatible with a disk-like source in a trans-Neptune comet belt, which they call the "Kuiper Belt". A possible member of the Kuiper Belt was first discovered in 1992 at a distance of 41 AU from the Sun (1992 $QB_1$, reported in Jewitt & Luu 1993), and several additional discoveries have been reported since (Jewitt & Luu 1995).

The dynamical structure of this putative comet population as determined by the long-term (conservative) gravitational perturbations by the planets has been the subject of two recent studies (Levison & Duncan 1993, Holman & Wisdom 1993). These studies assumed a uniformly distributed initial population in near-circular, low-inclination orbits, and the planets in their present orbital configuration; they sought to determine the extent and nature of any orbital instabilities that might operate on billion year timescales to transport those putative comets into planet-crossing orbits. However, any trans-Neptune population of planetesimals was undoubtedly subject to dynamical evolution during the planet formation process, and the initial conditions assumed in the above studies are not necessarily representative of the state of the Kuiper Belt at the end of planet formation, as acknowledged in Holman & Wisdom 1993. Here I discuss the "dynamical sculpting" of the Kuiper Belt that would have occurred due to the early orbital evolution of the outer planets (during the late stages of their formation) as predicted by the "resonance capture theory" for the origin of Pluto's orbit. The results of this study indicate that the Kuiper Belt would have been "sculpted" into a highly non-uniform distribution early in Solar System history, and this structure would be largely preserved to the present epoch: the region beyond Neptune's orbit and up to approximately 50 AU heliocentric distance should have most of the primordial small bodies locked in orbital resonances with Neptune, particularly the 3:2 and the 2:1 orbital resonances which are located at semimajor axes of approximately 39.4 AU and 47.8 AU, respectively.

The rest of this paper is organized as follows. In section 2, I summarize the "resonance capture theory" for the origin of Pluto's orbit. Sections 3 and 4 describe the numerical simulations conducted to determine the implications of this theory for the dynamical structure of the trans-Neptunian Solar system. Section 5 provides a discussion of the results, a comparison with previous theoretical studies of the Kuiper Belt, and some consequences for observational surveys of the outer Solar system.

## 2. Resonance capture theory for the origin of Pluto's orbit

The phenomenon of capture into resonance as a result of some slow dissipative forces is common in nature, and there exists a large body of literature devoted to its study. A well-developed Solar system example is the formation of orbit-orbit resonances amongst the satellites of the giant planets by the action of slow tidal dissipation (see Peale 1986 and Malhotra 1994a for reviews). In general, capture into a stable (long-lived) orbit-orbit resonance is possible when the orbits of two bodies approach each other as a result of the action of some dissipative process. In Paper I, I proposed that Pluto may have been captured into the 3:2 resonance with Neptune during the late stages of planet formation, when Neptune's orbit expanded outward as a result of angular momentum exchange with residual planetesimal debris. The physics of this mechanism is summarized below.

The giant planets' gravitational perturbations were instrumental in clearing their inter-planetary regions of the residual unaccreted planetesimal debris. It is believed that the Oort Cloud — which is a roughly isotropic distribution of comets surrounding the planetary system at distances in excess of $\sim 10^4$ AU – was populated by icy planetesimals scattered outward from the vicinity of the giant planets; estimates of the total mass of the Oort Cloud are in the range $10 - 10^2 M_\oplus$, based upon observations of long period comets and extensive theoretical modeling (cf. Weissman 1990). While the formation and dynamical evolution of the Oort Cloud has been a subject of extensive research (indeed, it is a sub-specialty within planetary science), relatively little attention has been given to the back reaction on the planets themselves of the planetesimal scattering process that populated the Oort Cloud.

Consider the scattering of a planetesimal of mass $m_c$ by a planet of mass $M$ at orbital radius $a$. If the planetesimal is initially in a near-circular orbit similar to that of the planet and is ejected to a Solar system escape orbit, it follows from conservation of angular momentum that the planet suffers a loss of orbital angular momentum and a corresponding change of orbital radius, $\delta a$, given by[1]

$$-\frac{\delta a}{a} \simeq \frac{m_c}{M} \qquad (1)$$

For a planetesimal scattered outward, but one that does not achieve a Solar system escape orbit (remaining bound, for example, in the Oort Cloud), the numerical coefficient in the right hand side of Eqn. (1) would be slightly smaller than unity. Conversely, planetesimals scattered inward would cause an increase of orbital radius and angular momentum of the planet. A single, massive planet scattering a population of planetesimals in near-circular orbits in the vicinity of its own orbit would, to first order, suffer no net change of orbital radius as it scatters approximately equal numbers of planetesimals inward and outward.

However, the four Jovian planets acting together evolve differently from this simple picture, as first pointed out by Fernandez & Ip 1984 who modeled the late stages of accretion of planetesimals ("protocomets") by the proto-giant planets (and the concomitant exchange of energy and angular momentum between the planetesimals and the planets). Their numerical simulations showed a small decrease in orbital radius for Jupiter and significant increases in orbital radius for Saturn, Uranus and Neptune. The reason for this orbital evolution — in particular that of Neptune — may be understood by means of the following heuristic picture of the clearing of a planetesimal swarm from the vicinity of Neptune. Suppose that the mean specific angular momentum of the swarm is initially equal to that of Neptune. At first, a small fraction of planetesimals is accreted as a result of physical collisions, and of the remaining, there are approximately equal numbers of inward and outward scatterings. To first order, these cause no net change in Neptune's orbit. However, the subsequent fate of the inward and outward scattered planetesimals is not symmetrical. Most of the inwardly scattered objects enter the zones of influence of the inner Jovian planets (Uranus, Saturn and Jupiter). Of those objects scattered outward by Neptune, some are lifted into wide, Oort Cloud orbits while others return to be accreted or rescattered; a fraction of the latter is again (re)scattered inwards where the inner Jovian plan-ets control the dynamics. In particular, the massive Jupiter is very effective in causing a systematic loss of planetesimal mass by ejection into Solar System escape orbits. Therefore, as Jupiter preferentially removes the inward scattered Neptune planetesimals, the planetesimal population encountering Neptune at later times is increasingly biased towards objects with specific angular momentum (and energy) larger than Neptune's. Encounters with this planetesimal population produce effectively a negative drag on Neptune which results in Neptune experiencing a net gain of angular momentum and energy, hence an increase in its orbital radius. Note that Jupiter is, in effect, the source of this angular momentum and energy; however, owing to its much larger mass, its orbital radius decreases by only a small amount.

---

[1] The expression given in Eqn. (1) in Malhotra 1993a is incorrect in that the coefficient on the right hand side of the equation was underestimated by about 20%; I am indebted to S. Tremaine for calling my attention to this.

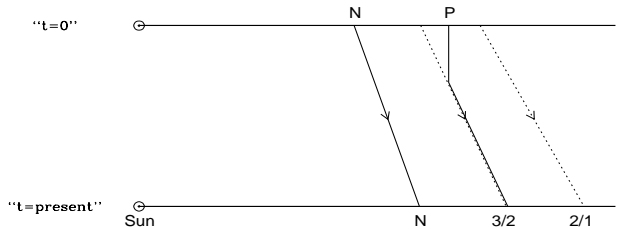

Fig. 1.— A schematic diagram to illustrate the outward radial migration of Neptune and its exterior orbital resonances during the late stages of planet formation. The distance from the Sun is along the horizontal direction. Neptune's outward orbital migration is shown along the path marked N—N. For clarity, only two first order resonances (3:2 and 2:1) are shown (dotted lines). A "Pluto" in an initially circular, non-resonant orbit beyond Neptune could have been captured into the 3:2 resonance and would evolve along the solid line path indicated by P—3/2.

If the above phenomenon did occur in the late stages of planet formation, one consequence of Neptune's orbital expansion is that its orbital resonances would have swept across a range of heliocentric distances comparable to its radial migration (see Figure 1). During this "resonance sweeping", a small body such as Pluto, initially in a near-circular orbit beyond Neptune, could have been captured and locked into an orbital resonance.

Resonance capture is a complicated dynamical process and a very active subject of research in the non-linear dynamics literature (see, for example, Wiggins 1991). The transition from a non-resonant to a resonant orbit depends sensitively upon initial conditions, and in the context of the Solar system, upon the nature of the resonant as well as non-resonant gravitational perturbations, and the rate of orbit evolution due to the dissipative effects. Under certain idealized conditions ("single resonance"), and in the limit of slow "adiabatic" orbit evolution, the probability of resonance capture is relatively straightforward to calculate (Henrard & Lemaitre 1983, Borderies & Goldreich 1984). Such a calculation shows that the capture probability for the 3:2 Neptune resonance is 100% for initial eccentricity (before the resonance encounter) less than $\sim 0.03$; the capture probability decreases monotonically for higher initial eccentricities: it is less than 10% for initial eccentricities exceeding 0.15.

Once an object was captured into an orbital resonance, perturbations from Neptune would transfer sufficient angular momentum to it to maintain it in the resonance by expanding its orbit in concert with that of Neptune. A by-product of this evolution would have been the rapid excitation of the object's orbital eccentricity. This is most readily seen in the following simplified analysis of the first-order perturbations of Neptune on a test-particle orbit. Close to an exterior $j+1:j$ orbital resonance, the first-order perturbation equations for the mean motion, $n$, and eccentricity, $e$, of the particle are (cf. Brouwer & Clemence 1961):

$$\dot{n} = 3(j+1)\mu_N n^2 e f(\alpha) \sin\phi,$$
$$\dot{e} = -\mu_N n f(\alpha) \sin\phi, \quad (2)$$

where $\mu_N = m_N/M_\odot$ is the mass of Neptune relative to the Sun, $\alpha = a_N/a < 1$ is the ratio of semimajor axes of Neptune and the test particle, $f(\alpha)$ is a positive function that can be expressed in terms of Laplace coefficients. $\phi = (j+1)\lambda - j\lambda_N - \varpi$ is the critical resonance angle, with $\lambda$ and $\varpi$ the mean longitude and the longitude of perihelion, respectively, of the test particle, and $\lambda_N$ the mean longitude of Neptune.

If the test particle is captured into resonance, its mean motion becomes locked to that of Neptune, so that the following conditions hold:

$$(j+1)n \simeq j n_N,$$
$$(j+1)\langle\dot{n}\rangle \simeq j\langle\dot{n}_N\rangle, \quad (3)$$

where $\langle \dot{n}_N \rangle$ is the rate of change of Neptune's mean motion as its orbit expands. It then follows from equations (2) and (3) that

$$\langle \frac{de^2}{dt} \rangle \simeq -\frac{2}{3(j+1)} \langle \frac{\dot{n}_N}{n_N} \rangle = \frac{1}{(j+1)} \langle \frac{\dot{a}_N}{a_N} \rangle, \quad (4)$$

where the last equality follows from the Keplerian relation between mean motion and semimajor axis ($n^2 a^3 = constant$). Therefore, upon capture into resonance, the test particle's eccentricity is pumped up at a rate determined by the average rate of expansion of Neptune's orbit. The previous equation can be integrated to yield

$$e_{\text{final}}^2 \simeq e_{\text{initial}}^2 + \frac{1}{j+1} \ln \frac{a_{N,\text{final}}}{a_{N,\text{initial}}} \quad (5)$$

where $a_{N,\text{initial}}$ refers to the value of Neptune's semimajor axis at the point of resonance capture. Note that for initially near-circular orbits, the final eccentricity is insensitive to the initial eccentricity and depends only upon the extent of orbital migration of Neptune.

Applying this result to Pluto, one concludes that if an initially near-circular Pluto was captured into the 3:2 resonance ($j = 2$) with Neptune and its eccentricity was resonantly pumped up to its current value of 0.25 in the subsequent evolution, then Neptune was at $a_N \approx 25$ AU at the point of resonance capture. Pluto's initial orbital radius can then be inferred to have been near 33 AU. The 5 AU outward radial migration inferred for Neptune must be regarded as a lower limit, as $a_N \simeq 25$ AU is the stage at which Pluto is inferred to have been captured in the 3:2 Neptune resonance. Other trans-Neptune objects captured in the 3:2 resonance at earlier or later times would have their eccentricities pumped up to larger or smaller values, respectively, than that of Pluto. Other first order resonances of importance in trapping trans-Neptune objects are the 4:3 and the 2:1, currently located at semimajor axes of 36.5AU and 47.9AU, respectively. The 5:3 second-order resonance located at 42.3 AU also has a significant capture probability, and capture is also possible in other higher order resonances such as the 7:5, and 7:4 located at 37.7 AU and 43.7 AU, respectively.

## 3. Numerical Model for the evolution of the trans-Neptune population

In the late stages of planetary accumulation, the exact magnitude of the radial migration of the Jovian planets due to their interactions with residual planetesimals is difficult to estimate without a full-scale N-body model. The work of Fernandez & Ip 1984 is suggestive, but simulations of this process stand to profit by the refinements in computational technology that have occurred in the last ten years. For example, the limitations of the software and hardware available to Fernandez & Ip limited the total number of bodies in their simulations to about 2000; therefore, in order to start with a reasonable total initial mass in the planetesimal disk, the masses of their individual planetesimals were in the rather exaggerated range of $(0.02 - 0.3) M_\oplus$. Another approximation in their model is the neglect of all but very close encounters between the (proto-) giant planets and the planetesimals. I expect to improve upon this in future work.

For the present, I do not attempt such detailed modeling here, but rather continue to use the simple model outlined in Paper I. Accordingly, the system consists of the Sun and the four Jovian planets with their present masses, together with a population of massless "test particles" representing trans-Neptunian objects. The massive planets perceive their full mutual gravitational interactions and also perturb the test particles, but the latter are non-interacting and do not perturb the planets. (It is perhaps worth emphasizing that there is little overlap between the *trans-Neptunian* population of small objects considered here and the population of residual planetesimals in the *immediate vicinity* of the giant planets whose gravitational interactions with the planets are presumed to be driving the radial migration of those planets.) The orbital migration of the Jovian planets is modeled by a time variation of their semimajor axes according to the following prescription:

$$a(t) = a_f - \Delta a \exp(-t/\tau), \quad (6)$$

where $a_f$ is the semimajor axis at the current epoch and $a_i \equiv a_f - \Delta a$ is the semimajor axis at the starting point ($t = 0$) of the simulation. (The epoch "$t = 0$" refers to a time in the late stages of the genesis of the Solar system when the formation of the gas giant planets was largely complete, the Solar Nebula had lost its gaseous component, and the subsequent evolution was dominated by gravitational interactions amongst the planets and the residual planetesimals. See Levy & Lunine 1993 for details on the various stages of Solar system formation.) The orbit evolution given by Eqn. (6) was implemented in the equations of motion by means of an additional "drag" force

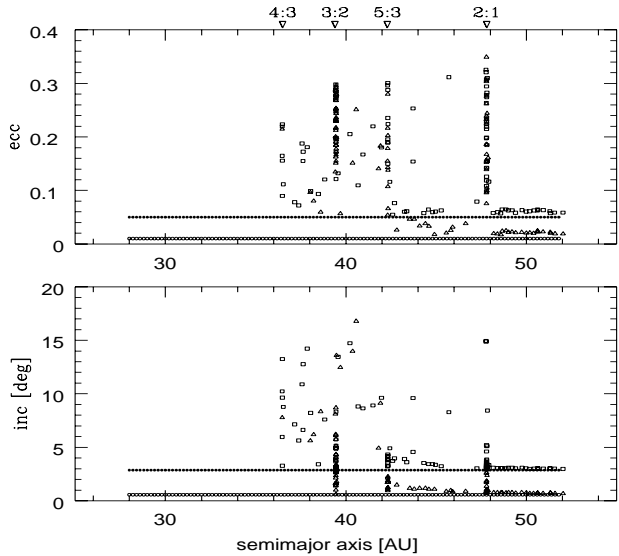

Fig. 2.— The initial and final orbits of the surviving Kuiper Belt objects in Run 1 ("thin disk": initial $e = i = 0.01$ shown by open circles; final elements shown by triangles) and Run 2 ("thick disk": initial $e = i = 0.05$ shown by solid circles; final elements shown by squares). For clarity, I show the final semimajor axes averaged over the last 2 million yr period in the runs. The locations of several major orbital resonances with Neptune are indicated at the top of the upper panel. The concentrations of objects at these resonances have significantly large eccentricities and inclinations.

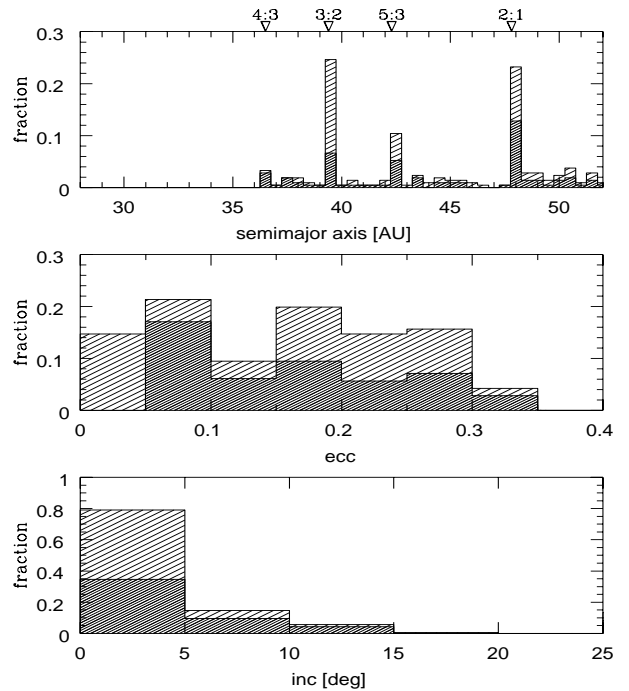

Fig. 3.— The distribution of orbital semimajor axes, eccentricities and inclinations of the surviving Kuiper Belt objects in Runs 1 (lighter shading) and 2 (darker shading). A large fraction of the objects beyond Neptune and up to $a \approx 50$ AU are found in narrow resonance regions, as indicated in the top panel. (The distributions for Runs 1 and 2 are quite similar).

on each planet along the direction, $\hat{\mathbf{v}}$, of the orbital velocity given by

$$\Delta \ddot{\mathbf{r}} = \frac{\hat{\mathbf{v}}}{\tau} \left\{ \sqrt{\frac{GM_\odot}{a_f}} - \sqrt{\frac{GM_\odot}{a_i}} \right\} \exp(-\frac{t}{\tau}) \quad (7)$$

The numerical method described in Malhotra 1994b was used for the orbit integrations; this method is based upon a second order symplectic map (see Wisdom & Holman 1991) but modified for additional nongravitational forces. Note that the integration follows the orbits of the four major planets self-consistently (i.e. their mutual gravitational interactions are fully accounted for even as their orbits expand).

4. **Results**

Here I report the results of several runs based upon the model described above. In all cases, the parameters for the planetary orbit evolution were as follows. The $\Delta a$ (cf. Eqn. 6) were chosen to be $-0.2, 0.8, 3.0$ and $7.0$ AU for Jupiter, Saturn, Uranus and Neptune, so that the initial semimajor axes of these planets were approximately 5.4, 8.7, 16.3 and 23.2 AU, respectively. These are within the ranges of radial displacements of these planets found in the Fernandez & Ip 1984 calculations. The planet masses and other initial orbital elements were taken from Nobili et al. 1989.

The first two runs described here were aimed at determining the current state of a primordial population of small objects beyond Neptune in the Kuiper Belt (up to approximately 50 AU heliocentric distance). In each of these runs, there were 120 test particles – representing the Kuiper Belt objects – with initial semimajor axes distributed uniformly in the range 28–52 AU and all angles (longitude of perihelion, longitude of ascending node and mean longitude) chosen randomly from the range $(0, 2\pi)$. In Run 1 (a "thin disk"), the initial eccentricities and inclinations were set to 0.01; in Run 2 (a "thicker disk"), the initial eccentricities and inclinations were set to 0.05. The timescale $\tau$ for the radial migration of the planets was taken to be $2 \times 10^6$ yr. The system was integrated for a period of $2 \times 10^7$ yr which is 10 times the assumed orbital evolution timescale, $\tau$. At the end of the integration, the final orbits of the planets are very similar to their presently observed orbits. The integration of those test particles that suffered close approaches, i.e. within a Hill sphere radius, with any planet was terminated: the object was presumed to have failed to survive as a Kuiper Belt object. It was found that all objects with initial semimajor axes less than 30 AU and a small fraction of those with initial semimajor axes between 30 AU and 34 AU failed to survive. The survival rate in the "thicker disk" population was slightly lower than in the "thinner disk" (85% vs. 91%). (In general, the survival rate can be expected to be sensitive also to the planetary orbital evolution timescale, $\tau$, as discussed below.) Figures 2 and 3 summarize the results obtained in these two runs. Noteworthy features of the surviving test particle population are: (i) the final semimajor axes are larger than 36 AU; (ii) the population is highly concentrated at primarily two resonances with Neptune – the 3:2 and the 2:1 resonances located near 39.4 AU and 47.8 AU, respectively; smaller concentrations are also found at the 5:3, 4:3, 7:5 and 7:4 resonances located at 42.3 AU, 36.5 AU, 37.7 AU and 43.7 AU, respectively; and (iii) the objects surviving in resonances have significant orbital eccentricities, typically 0.1–0.3 (thus their perihelion distances can be as small as $\sim 27$ AU).

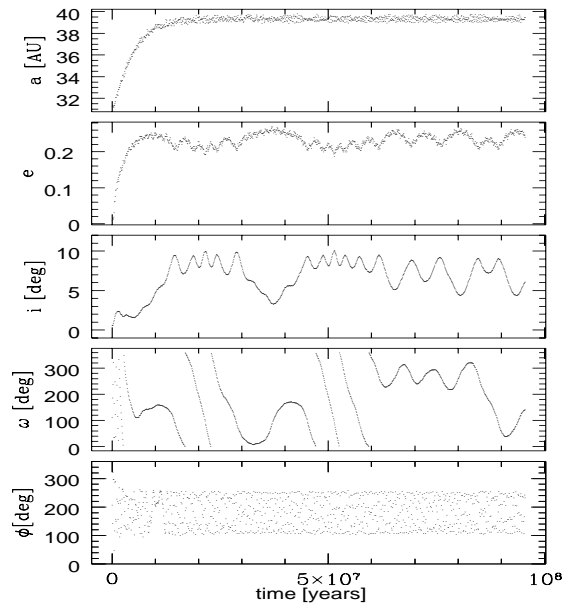

Fig. 4.— A typical example of the orbital evolution of a Pluto-like Kuiper Belt object captured in the 3:2 resonance with Neptune. The 3:2 resonance angle, $\phi = 3\lambda - 2\lambda_N - \varpi$, settles into stable librations about $180°$. There ensues a secular transfer of angular momentum from Neptune to the object that maintains it in the resonance; the object's orbital semimajor axis increases (in concert with Neptune's) and its eccentricity is pumped up (cf. Eqn. 2). In many cases, the inclination is also pumped up, but this depends sensitively upon the initial conditions of the orbit; the inclination behavior is highly correlated with the behavior of the argument of perihelion, $\omega$, which often exhibits long periods of libration about either $+90°$ or $-90°$.

Three other runs were made with a special focus on Pluto-like orbits captured in the 3:2 orbital resonance with Neptune. In each of these, 120 test particles with initially near-circular, co-planar orbits ($e = i = 0.01$) distributed uniformly in $a \in (29, 35) AU$ were integrated for a period of 100 million years with the same model described above. The three runs differed only in the timescale $\tau$ of orbit evolution of the planets; $\tau$ was set equal to 2, 4 and 10 million years in Run 3, Run 4 and Run 5, respectively. The survival rates in these runs were 79%, 68% and 53%, respectively. As expected by design, most of the surviving bodies were locked in the 3:2 resonance at the end of the evolution. (The exceptions were a few objects which were found locked in other nearby resonances.) An example of the typical orbital evolution of a test particle captured in the 3:2 resonance is shown in Figure 4 which displays the time variation of those variables that are of particular interest in the dynamics of Pluto's orbit. Observe that the semimajor axis stabilizes at the 3:2 resonance value; the eccentricity and inclination are both amplified; the 3:2 resonance angle, $\phi = 3\lambda - 2\lambda_N - \varpi$ (where $\lambda_N$ and $\lambda$ are the mean longitudes of Neptune and the test particle, and $\varpi$ is the longitude of perihelion of the test particle), settles into stable libration about $180°$, and the argument of perihelion, $\omega$, also exhibits librations. Long-lived librations of $\omega$ about either $+90°$ or $-90°$ (i.e. perihelion near the maxi-mum excursion above or below the mean plane of the Solar System) are found in approximately one-third of the objects trapped in the 3:2 Neptune resonance; often, the $\omega$ libration center hops between $+90°$ and $-90°$ over $10^7$ yr timescales (cf. Figure 4). The $\omega$ librations are strongly correlated with large perturbations of the inclination. (This suggests that the resonance represented by the argument-of-perihelion libration may be responsible for the inclination excitation. However, this conjecture must await further analysis.)

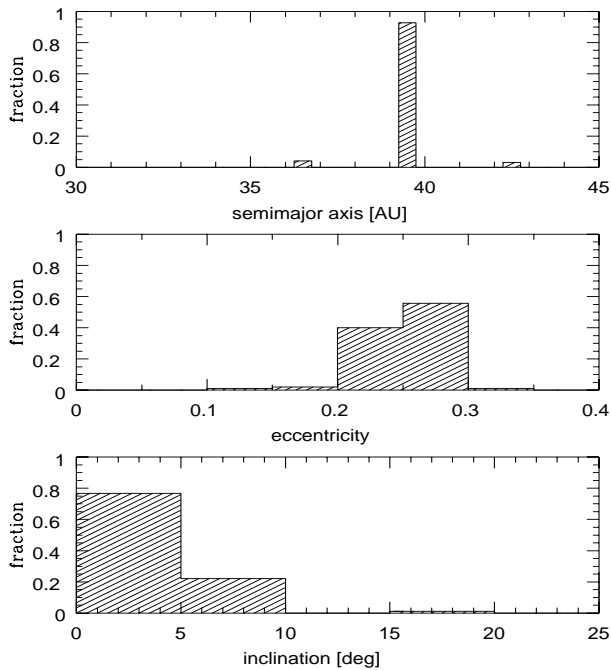

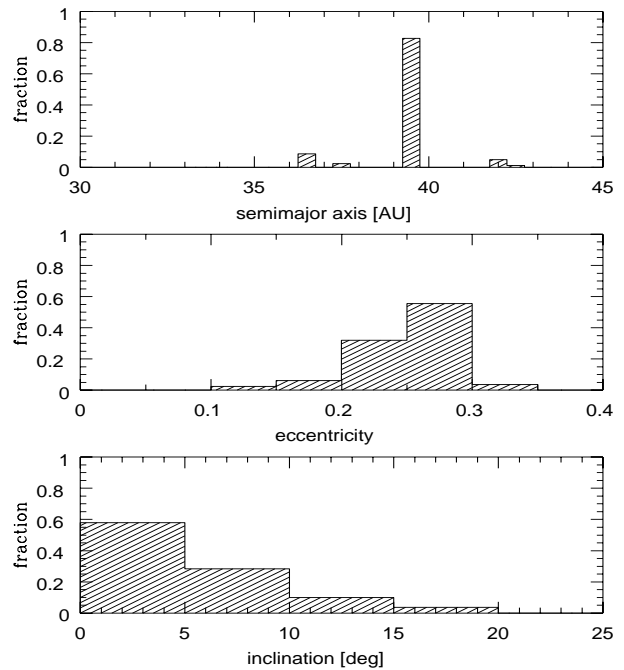

Fig. 5.— The final distribution of $a$, $e$ and $i$ in Run 3 in which the initial distribution of objects was in near-circular, co-planar orbits in an annulus with $a$ between 29 AU and 35 AU (designed to yield mostly Pluto-like orbits locked in the 3:2 orbital resonance with Neptune and with $e$ between 0.2 and 0.3). The timescale of orbital migration of the planets in this run was $\tau = 2 \times 10^6$ yr.

Fig. 6.— Same as Figure 5 but for Run 4 in which the timescale of orbital migration of the planets was $\tau = 4 \times 10^6$ yr.

The distributions of the orbital elements of the survivors in these three runs are shown in Figures 5–7. Noteworthy features are as follows. The eccentricity distribution is virtually identical in the three runs, with the vast majority of the objects having $e$ in the range 0.2 to 0.3 (as expected by design). Most of the objects remain in relatively low inclination ($< 10°$) orbits, but a small fraction (up to 10% in Run 7) have their inclinations pumped up to higher values (15-20 degrees), comparable to that of Pluto. There is a correlation between higher inclinations and larger $\tau$ (i.e. slower evolution rate for the planetary orbits).

## 5. Conclusions and Discussion

The unusual properties of Pluto's orbit may be a natural consequence — and a signature — of the early dynamical evolution in the outer Solar system. The studies presented here, together with those reported in previous work Malhotra (1993a,1995) provide support for this case. I had shown previously that if Neptune's orbit expanded during the late stages of planet formation, then Pluto could have been captured from an initially near-circular, low inclination non-resonant orbit beyond Neptune, into its current 3:2 resonant orbit; during this evolution, its eccentricity would have been pumped up to the observed Neptune-crossing value while maintaining long-term orbital stability. The large inclinations and the argument-of-perihelion librations found in the work presented here augment the previous results and enhance the plausibility of the resonance capture theory for the origin of Pluto's orbit. However, it is also clear that there is a rather small probability for obtaining inclinations as high as Pluto's.

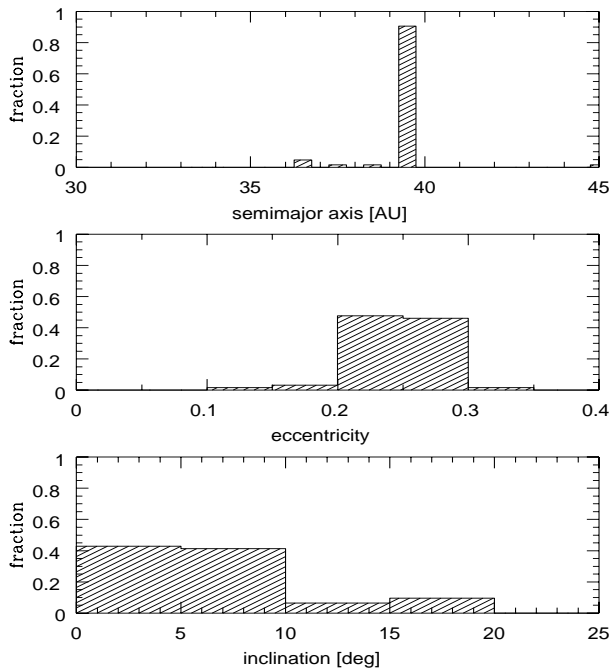

Fig. 7.— Same as Figures 5 and 6, but for Run 7 in which the timescale of orbital migration of the planets was $\tau = 1 \times 10^7$ yr. Note that there is little variation in the eccentricity distributions in Figures 5, 6 and 7, but the inclination distribution is progressively broader.

From the observed characteristics of Pluto's orbit, I have shown that it is possible to infer the "initial" locations of both Neptune and Pluto and thus obtain a lower limit of about 5 AU for the magnitude of Neptune's orbital expansion. (With further modeling, one can anticipate that it will be possible to also self-consistently infer the "initial" orbits of the other giant planets.) These inferences are of considerable import for many aspects of planet formation. Some examples are: the provenance of the outer planets may derive from a different and perhaps larger range of heliocentric distance than the immediate vicinity of their current orbits; Jupiter's orbital migration could have influenced the planetesimal dynamics and accumulation process in the region of the asteroid belt; the capture of irregular satellites may have been significantly aided by the radial migration of the (proto-)giant planets. However, a discussion of these points is beyond the scope of the present work.

In this paper, I have followed this "resonance capture" scenario for the origin of Pluto's orbit to its logical next step in studying its implications for the architecture of the Solar system beyond Neptune, i.e. the "Kuiper Belt" of comets approximately between Neptune's orbit and 50 AU. The numerical experiments reported here indicate that the dynamical structure of the Kuiper Belt is dominated by concentrations of objects trapped in orbital resonances with Neptune, particularly at the 3:2 and the 2:1 resonances. These resonant objects move on highly eccentric orbits, with a significant fraction on Neptune-crossing orbits; the inclinations of most of the objects remain low (less than 10°), but a small fraction (up to ∼ 10%) are in the 15-20 degree range. Libration of the argument-of-perihelion (about ±90°) is not an uncommon occurrence amongst the resonant objects.

The numerical results also show that the inclination distribution is sensitive to the rate of orbital evolution of the giant planets: longer timescale of the orbit evolution is correlated with higher inclinations (cf. the third panel in Figures 5, 6 and 7). The timescales used in the numerical experiments were chosen primarily for computational convenience, and ranged from 2 to 10 million years. As the libration period for the argument-of-perihelion libration (which is correlated with inclination excitation) is several million years, it is clear that the timescale of this resonance is comparable to the timescale of orbital evolution of the giant planets in these numerical experiments. Thus, the sensitivity of the inclination distribution to the rate of orbital evolution of the planets is not surprising as the evolution is not "adiabatic" on this timescale. It is worth noting that the numerical experiments of Fernandez & Ip 1984 indicate orbital evolution timescales of several tens of millions of years, and I expect that more realistic modeling in the future will help determine this timescale better. Comparison of these inclination distributions with observations (when statistically significant numbers of observations of trans-Neptunian objects and their orbital elements are available) may provide a diagnostic of the rate of radial migration of the planets during the late stages of their evolution.

Of course, the integrations reported here were only $2 \times 10^7 - 10^8$ years long, and one might ask if the subsequent nearly-conservative evolution over $10^9$ yr timescales might not change the profile of the surviving objects. I have extended the orbit integrations to 1 Byr in a few individual cases and have found that the resonance protection remained in place over this longer time period. Although suggestive, this is of course not proof of orbital stability over the 4.5 Byr age of the Solar system. A slow leakage of objects out of the orbital resonances over this timescale – either due to collisional evolution, or possibly due to long-term purely gravitational effects, or both – cannot be ruled out, and may be necessary to supply the short period comet population. In this context, it is also worthwhile to note that resonance capture is not 100% efficient: a small fraction of the original trans-Neptunian population survives the "resonance sweeping" in its primordial non-resonant low-eccentricity, low-inclination orbits, and may also contribute to the flux of short-period comets. A quantitative study of this point remains to be done.

*Comparison with previous theoretical studies of the Kuiper Belt*

There have been two previous theoretical investigations of the dynamics of small bodies in the outer Solar sytem. Holman & Wisdom 1993 and Levison & Duncan 1993 have studied test particle orbital stability on $10^9$ yr timescales. For test particles in nearly circular and low-inclination orbits beyond Neptune, these studies found orbital instability on short timescales ($< 10^7$ yr) interior to 33-34 AU, an intricate structure of interspersed regions of stability and instability in the semimajor axes range of 34 AU to 43 AU, and substantially stable orbits beyond 43 AU. The intricacy of the dynamical structure appears to be particularly acute near orbital resonance locations, but as these features are not analyzed in detail in either paper, it is difficult to draw clear conclusions about the relative number density of Kuiper Belt objects that might be expected in these regions at the present epoch. Holman & Wisdom also found substantial 'bumps' in the eccentricities and inclinations near the 3:2 resonance and in the eccentricity at the 2:1 resonance, indicating potential sources of planet-crossing cometary orbits in these zones. Both studies indicate that once an object became Neptune-crossing, it typically suffered a close encounter with the planet in short order (although a few exceptional cases of long-lived Neptune-crossers were detected).

In other words, Neptune-crossing orbits do not typically enjoy long-term stability due to resonance protection even when they originate close to resonance locations. This is not surprising, for the stable orbits near Neptune resonances exist only in exceedingly narrow zones in phase space, and a random sampling of initial conditions would have low probability of hitting these zones.

The resonance sweeping mechanism studied in the present paper predicts the inner edge of stability at approximately 36 AU in semimajor axis, and clear and strong enhancements in the stability and relative population density of objects in orbital resonances with Neptune. In effect, the extensive non-resonant regions of space are depleted of objects as those objects are swept into resonant orbits. In spite of the low phase space volume represented by the resonance libration regions, the dissipative nature of the dynamical evolution in this model very strongly enhances the occupancy rate of the resonance zones. This model predicts somewhat greater "dynamical erosion" in the Kuiper Belt than the above-mentioned studies of Holman & Wisdom and Levison & Duncan, particularly in the regions in-between the resonances. The high eccentricities of the resonance-trapped objects make a large fraction of them Neptune-crossing yet dynamically long-lived.

*Implications for observational surveys of the Kuiper Belt*

The chief dynamical feature of the resonance-locked orbits is that the longitude of perihelion *avoids* Neptune. (This fact can be inferred from the librations of the resonance angle.) For example, in the case of the 3:2 resonance, the resonance angle librates about $180°$, and its maximum libration amplitude (for stable orbits) is about $90°$; the longitude of perihelion librates about a center $\pm 90°$ away from Neptune's mean longitude. The situation is more complicated for the 2:1 resonance, as the location of the center of libration is a strong function of the orbital eccentricity; the center of the perihelion oscillation (relative to Neptune's mean longitude) is at $180°$ for very small eccentricities; for larger eccentricities (exceeding $\sim 0.03$), the libration center bifurcates and the two new libration centers drift away (symmetrically) from $180°$ to nearly $+45°$ and $-45°$ for eccentricities near 0.3. These properties of the resonant orbits will be discussed in further detail in a future publication

|    | 3:2 resonance | | 2:1 resonance | |
| --- | --- | --- | --- | --- |
| $m$ | $s$(peri) | $s$(aph) | $s$(peri) | $s$(aph) |
| 17 | 690 | 1570 | 1020 | 2324 |
| 20 | 173 | 395 | 256 | 584 |
| 22 | 69 | 157 | 102 | 232 |
| 25 | 17 | 39 | 26 | 58 |

Table 1: The first column is magnitude, $m$; $s$(peri) and $s$(aph) are the minimum radii (in km) of objects of magnitude less than $m$ at perihelion and aphelion, respectively, in orbits of eccentricity 0.2. Columns 2 and 3 are for objects in the 3:2 Neptune resonance (semimajor axis = 39.4 AU); columns 4 and 5 are for objects in the 2:1 Neptune resonance (semimajor axis = 47.8 AU). We assume a mean geometric albedo $A = 0.1$.

Malhotra (1995, in preparation).

If most Kuiper Belt objects do indeed exist primarily in two resonance bands (the 3:2 and the 2:1), are there particular observational strategies that might improve the discovery rate of these objects? As the results in section 4 indicate, the inclinations of these objects do not exceed 20 degrees (more than 80% of them remain less than 10 degrees). Therefore, sky surveys within $\pm 20°$ of the ecliptic are likely to yield most of the detectable objects.[2] In a magnitude-limited sky survey, the minimum size of a detectable outer Solar system object will vary with its distance and its albedo. Assuming that observations are made at opposition (geocentric distance = heliocentric distance − 1 AU), the size-magnitude-heliocentric distance relationship is given by

$$s = 1803(\frac{A}{0.1})^{-1/2} r(r-1) 10^{-0.2m} \qquad (8)$$

where $s$ is the object radius in km, $A$ is the geometric albedo, $r$ is the heliocentric distance in AU, and $m$ is the magnitude. For a typical orbital eccentricity of $\sim 0.2$, the perihelion to aphelion distance varies from 31 AU to 47 AU in the 3:2 resonance, and from 38 AU to 58 AU in the 2:1 resonance. The minimum radii of detectable objects at perihelion and aphelion in these resonant orbits are listed in Table 1 for several limiting magnitudes. For given magnitude and albedo, the minimum radius of a detectable object varies by a factor of approximately $\sim (1+e)^2/(1-e)^2$ from perihelion to aphelion. For $e = 0.2$, this factor is $\sim 2.3$. For illustrative purposes, let us assume that the Kuiper Belt objects have a power law size distribution, $n(s)ds \propto s^{-q} ds$, with index $q = 2$ (cf. Jewitt & Luu, 1995). Then, the number of resonance-trapped objects that are potentially detectable at perihelion is a factor $\sim 2.3$ greater than those detectable at aphelion. Recall that the perihelia of the 3:2 resonant objects librate near $\pm 90°$ from Neptune. The 2:1 resonant objects (with $e \sim 0.2$) have perihelia librating near $+60°$ or $-60°$ from Neptune. Assuming that, at any given epoch, the resonant objects are uniformly distributed in ecliptic longitude, those objects at longitudes defined by the two quadrants $(45°, 135°)$ and $(-135°, -45°)$ relative to Neptune would be at or close to perihelion. Therefore, without going through a detailed statistical analysis (to take account of factors such as the distributions in eccentricity and resonance libration amplitude), a crude estimate is that discovery rates in these regions of the sky would be greater by a factor of about 2 compared to the two quadrants outside these ranges of longitude. With Neptune currently near $290°$ ecliptic longitude, the favored quadrants correspond to ecliptic longitude ranges $(335°, 65°)$ and $(155°, 245°)$.

It is interesting to consider the detectability of a Pluto-like object in the 2:1 resonance with Neptune. Such an object (with radius $s \sim 1000$ km, mean geometric albedo $A \sim 0.5$, and orbital eccentricity 0.25) would vary in brightness from $m \simeq 15.0$ to $m \simeq 17.3$ from perihelion to aphelion. Of all the sky surveys for outer Solar system objects to date, Tombaugh's (1961) search is the most likely to have detected such a body. This survey covered all longitudes within approximately $\pm 15°$ of the ecliptic, with a limiting mag-nitude of about 16.5. Thus, a Pluto-like object in the 2:1 resonance would have been detected if it were at a heliocentric distance less than 49 AU. In a 2:1 resonant orbit (semimajor axis = 47.8 AU) with $e = 0.25$, an object spends only about 35% of its orbital period at heliocentric distances less than 49 AU. Therefore, in Tombaugh's survey, there was a $\sim 35\%$ chance of detecting such an object. The more recent surveys for outer Solar system objects (Kowal 1989, Luu & Jewitt 1988, Levison & Duncan 1990, Jewitt & Luu 1995) all had limiting magnitudes exceeding 17.3, but they also had much smaller sky coverage; detection probability is much smaller in these surveys for the latter reason. Therefore, a Pluto-like object in the 2:1 resonance cannot yet be ruled out.

---

[2] It should be noted that about a third of the resonant trapped objects also exhibit argument-of-perihelion libration; this means that at perihelion (when they are brightest and thus most likely to be detected), such objects would also be near their greatest ecliptic latitude (either above or below the ecliptic); observational searches that are too narrowly confined to the ecliptic may not detect these objects.

Observational surveys of the outer planetary system have recently reported the detection of several objects $\sim 100$ km in size which are possibly the larger members of the Kuiper Belt (Jewitt & Luu 1995), and it appears likely that even greater numbers of detections will be forthcoming in the near future. As the present work was in progress, Marsden (1994a,b) has reported on the possibility that several of the newly discovered objects may be in Pluto-like orbits, locked in the 3:2 resonance with Neptune. If this is confirmed, it would provide further corroboration for the "resonance sweeping" scenario. It is my hope that the present paper will contribute to the acceleration of these observational detections and their interpretation within models of the formation and evolution of the Solar system.

I am grateful to Scott Tremaine for a critical reading of this paper. This research was done while the author was a Staff Scientist at the Lunar and Planetary Institute which is operated by the Universities Space Research Association under contract no. NASW–4574 with the National Aeronautics and Space Administration. This paper is Lunar and Planetary Institute Contribution no. xxx.

**REFERENCES**

Applegate, J.H., M.R. Douglas, Y. Gursel, G. Sussman and J. Wisdom 1986, AJ, 92:176–194

Bailey, M.E. 1983, Nature, 302:399-400

Borderies, N. and P. Goldreich 1984, Cel. Mech., 32:127-136

D. Brouwer and G.M. Clemence 1961. Academic Press, New York

Burns, J.A., L.E. Schaffer, R.J. Greenberg and M.R. Showalter 1985, Nature, 337: 340-343

Cohen, C.J. and E.C. Hubbard 1965, AJ, 70:10-13

Dormand, J.R. and M.M. Woolfson 1980, MNRAS, 193: 171-174

Duncan, M., T. Quinn and S. Tremaine 1988, ApJ, 328:L69-L73

Farinella, P., A. Milani, A.M. Nobili and G.B. Valsecchi 1979, The Moon & the Planets 20:415-421

Fernandez, J.A. 1980, MNRAS, 192:481-491

Fernandez, J.A. and W.H. Ip. 1983, Icarus, 54:377-387

Fernandez, J.A. and W.H. Ip 1984, Icarus, 58:109-120

Hamid, S.E., B.G. Marsden and F.L. Whipple 1968, AJ73:727-729

Harrington, R.S. and T.C. van Flandern 1979, Icarus, 39:131-136

Henrard, J. and A. Lemaitre 1983, Cel. Mech., 30:197-218

Holman, M.J. and J. Wisdom 1993, AJ, 105:1987-1999

Jewitt, D.C. and J.X. Luu 1993, Nature, 362(6422):730-732

Jewitt, D.C. and J.X. Luu 1995, AJ, (in press)

Kowal, C.T. 1989, Icarus, 77:118-123

Kuiper, G.P. 1951, in *Astrophysics: A topical symposium*, J.A. Hynek, ed., McGraw-Hill, New York, p. 357-424

Levison, H.F. and M.J. Duncan 1990, AJ, 100:1669-1675

Levison, H.F. and M.J. Duncan 1993, ApJ, 406:L35-L38

Levison, H.F. and S.A. Stern 1994, Icarus, submitted.

Levy, E.H. and J.I. Lunine 1993, eds, *Protostars & Planets III*, The Univ. of Arizona Press, Tucson

Luu, J.X. and D. Jewitt 1988, AJ, 95:1256-1262

Lyttleton, R.A. 1936, MNRAS, 97:108-115

Malhotra, R. 1993a, Nature, 365:819-821 (Paper I)

Malhotra, R. 1994a, Physica D, 77:289-304. (Special issue on "Modeling the Forces of nature")

Malhotra, R. 1994b, Cel. Mech.&Dyn. Ast., 60:373-385

Malhotra, R. 1995, LPSC, XXVI:887-888

Malhotra, R. and J.G. Williams 1994, in *Pluto & Charon*, D. Tholen & S.A. Stern, eds., Univ. of Arizona Press, Tucson. (in press)

B.G. Marsden 1994a, IAU Circular 5983

B.G. Marsden 1994b, IAU Circular 6076

W.B. McKinnon and S. Mueller 1988, Nature, 335:240-243

McKinnon, W.B. and E.M. Parmentier 1986, in *Satellites*, J.A. Burns & M.S. Matthews, eds. Univ. of Arizona Press, Tucson. p. 718–763

Milani, A., A.M. Nobili and M. Carpino 1989, Icarus, 82:200-217

Nobili, A.M., A. Milani and M. Carpino 1989, A&A, 210:313-336

Olsson-Steel, D.I. 1988, A&A, 195:327-330

S.J. Peale 1986, in *Satellites*, J. Burns & M. Matthews, eds. Univ. of Arizona Press, Tucson. p. 159-223

Quinn, T.R., S. Tremaine and M.J. Duncan 1990, ApJ, 355:667-679

Standish, E. M. 1993, AJ, 105:2000-2006

Sussman, G.J. and J. Wisdom 1988, Science, 241:433-437

G. Tancredi and J.A. Fernandez 1991, Icarus, 93:298-315

Tombaugh, C.W. 1961, in *Planets and Satellites*, G.P. Kuiper & B.M. Middlehurst, eds. Univ. Chicago Press, Chicago. pp. 12-30

Weissman, P.R. 1990, Nature, 344:825-830

Whipple, F.L. 1964, Proc. Nat. Acad. Sci. 51:711-718

S. Wiggins 1991, Chaotic Transport in Dynamical Systems, Springer-Verlag, New York

Williams, J.G. and G.S. Benson 1971, AJ, 76:167-177

Wisdom, J. and M.J. Holman 1991, AJ, 102:1528-1538